\documentclass[preprint,superscriptaddress,showpacs]{revtex4}
\usepackage{epsfig,amsmath,graphicx,amssymb}
\usepackage[colorlinks=fals,linkcolor=blue]{hyperref}
\usepackage{amsfonts}
\usepackage{subfigure}
\usepackage{newlfont}
\usepackage{algorithm}
\usepackage{algorithmic}
\usepackage{multirow}
\usepackage{xcolor}
\usepackage{graphicx}

\newtheorem{lemma}{Lemma}

\def\be{\begin{equation}}
\def\ee{\end{equation}}
\def\bea{\begin{eqnarray}}
\def\eea{\end{eqnarray}}

\begin{document}
\title{Potts model based on a Markov process computation solves the community structure problem more theoretically and effectively }

\author{Hui-Jia Li}
\address{School of Management Science and Engineering, Central University of Finance and Economics, Beijing, 100081, PR China.}
\address{Academy of Mathematics and Systems Science, Chinese academy of Science, Beijing, 100190, PR China.}
\author{Yong Wang}
\address{Academy of Mathematics and Systems Science, Chinese academy of Science, Beijing, 100190, PR China.}
\address{National Center for Mathematics and Interdisciplinary Sciences, Chinese Academy of Sciences, Beijing 100190, China.}
\author{Ling-Yun Wu}
\address{Academy of Mathematics and Systems Science, Chinese academy of Science, Beijing, 100190, PR China.}
\address{National Center for Mathematics and Interdisciplinary Sciences, Chinese Academy of Sciences, Beijing 100190, China.}
\author{Junhua Zhang}
\address{Academy of Mathematics and Systems Science, Chinese academy of Science, Beijing, 100190, PR China.}
\address{National Center for Mathematics and Interdisciplinary Sciences, Chinese Academy of Sciences, Beijing 100190, China.}
\address{Key Laboratory of Random Complex Structures and Data Science, Chinese Academy of Sciences, Beijing 100190, China.}
\author{Xiang-Sun Zhang}
\email{zxs@amt.ac.cn}
\address{Academy of Mathematics and Systems Science, Chinese academy of Science, Beijing, 100190, PR China.}
\address{National Center for Mathematics and Interdisciplinary Sciences, Chinese Academy of Sciences, Beijing 100190, China.}
\date{\today}

\begin{abstract}
Potts model is a powerful tool to uncover community
structure in complex networks. Here, we propose a new
framework to reveal the optimal number of communities and stability of network structure by quantitatively analyzing the dynamics of Potts model.
Specifically we model the community
structure detection Potts procedure by a Markov process, which has a clear
mathematical explanation. Then we show that the
local uniform behavior of spin values across multiple timescales in the representation of the Markov variables
could naturally reveal the network's hierarchical community structure.
In addition, critical topological information regarding to
multivariate spin configuration could also be inferred from the spectral
signatures of the Markov process. Finally an algorithm is developed to determine fuzzy communities based on the optimal number of communities and the stability across multiple timescales. The effectiveness and
efficiency of our algorithm are theoretically analyzed as well as
experimentally validated.
\end{abstract}

\maketitle

%%%%%%%%%%%%%%%%%%%%%%%%%%%%%%%%%%%%%%%%%%%%%%%%%%%%%

\section{ Introduction}

Community structure detection \cite{Newman,Newman01, Newman02} is a main focus of
complex network studies. It has attracted a great deal of
attention from various scientific fields. Intuitively, community refers
to a group of nodes in the network that are more densely connected
internally than with the rest of the network. In the early stage,
these studies were restricted to the regular networks. Recently,
inspired by several common characteristics of real
networks\cite{Barabasi}, for example the scale-free property, the
majority of the studies focus on networks with practical applications.
In this meaning, community structure may provide insight into the relation between
the topology and the function of real networks and can be of
considerable use in many fields.

A well known exploration for this problem is via the modularity concept, which
is proposed by Newman et al. \cite{Newman,Newman01, Newman02} to quantify a network's partition. Optimizing modularity is effective for community
structure detection and has been widely used in many real networks.
However, as pointed out by Fortunato et al\cite{Fortunato},
modularity suffers from the resolution limit problem which concerns
about the reliability of the communities detected through the
optimization of modularity. In \cite{Arenas}, the authors claimed
that the resolution limit problem is attributable to the coexistence
of multiple scale descriptions of the network's topological structure,
while only one scale is obtained through directly
optimizing the modularity. In addition, the definition of modularity
only considers the significance of the link density from the static
topological structure, and it is unclear how the
modularity concept based community structure is correlated with the dynamics
behavior in the network.

Complementary to modularity concept, many efforts are devoted to understanding the
properties of the dynamical processes taken place in the underlying
networks. Specifically, researchers have begun to investigate the
correlation between the community structure and the
dynamics in networks. For example, Arenas et al. pointed out that the
synchronization reveals the topological scale in complex
networks\cite{Arenas1}. In addition, the Markov process on a network
was also extensively studied and used to uncover community structure
of the network\cite{Delvenne}--\cite{Zhou}. In
\cite{Weinan}\cite{Rosvall}, the Markov process on a network is
introduced to define the distances among network nodes, and an
algorithm is then proposed to partition the
network into communities based on these distances. In \cite{Delvenne}, the authors proposed to quantify and rank the network partitions in terms
of their stability, defined as the clustered autocovariance in the
Markov process taken place on the network.

Potts dynamical model has also been applied to uncover community
structure in networks. Detecting community by using Potts
model\cite{Wu}, also known as the superparamagnetic clustering
method, has been intensively studied since its
introduction by Blatt et al\cite{Blatt}. In the model, the Potts spin
variables are assigned to nodes of a network with community structure,
and the interactions exist between neighboring spins. Then the structural
clusters could be recovered by clustering similar spins
in the system, which have more interactions inside
communities than outside. The physical aspects of the method, such as its
dependence on the definition of the neighbors, type of interactions,
number of possible states, and size of the dataset, have been well
studied\cite{Wiseman}\cite{Agrawal}\cite{Ott}. Reichardt and
Bornholdt\cite{Reichardt} introduced a new spin glass Hamiltonian
with a global diversity constraint to identify proper community
structures in complex networks. The method allows one to identify
communities by mapping the graph onto a zero-temperature $q$-state Potts
model with nearest-neighbor interactions. Recently, Li et al\cite{Our1} noticed that a lot of
useful information related to community structure can be revealed by Potts model and the spectral
characterization. Despite those excellent works, uncovering the dynamic of spin configure across multiple timescales is still
a tough task and not yet been clearly answered. In essence, one can consider the time scale as an
intrinsic resolution parameter for the partition: over short time scales from the beginning, many
small clusters should be coherent; on the other hand as time evolves, there will be fewer and larger
communities that are persistent under the dynamic of Potts model. We need to measure the change of the stability or robustness\cite{Delvenne}
of spin configure as time evolves and furthermore find some reasonable partitions at intermediate timescales.
However, using Potts model alone is difficult to solve this problem.

We notice that the dynamics of Markov process can naturally reflect
the intrinsic properties of spin dynamics with community structures and
exhibit local uniform behaviors. However, the relationship between dynamics
of Potts model and the Markov process, has not been well studied.
In this work, using the Potts model and spin-spin correlation, we first
investigate this phenomenon, and then uncover the relation between
community structure of a network and its meta-stability of spin
dynamics, and further propose the signature of communities to
characterize and analyze the underlying spin configuration. For any
given network, one can straightforwardly derive critical information
related to its community structure, such as the stability of its community
structures and the optimal number of communities across multiple
timescales without using particular algorithms.
It overcomes the inefficiency of the classic methods, such as the resolution
limitation of Modularity $Q$ \cite{Fortunato}\cite{X}. Based on the
theoretical analysis, we then develop a parameter free
algorithm to numerically detect community structure, which is able
to identify fuzzy communities with overlapping nodes by associating each node with a
participation index that describes node's involvement in each
community. We also demonstrate that the algorithm is scalable and
effective for real large scale networks.

The outline of the paper is as follows. Section II introduces the
Potts model and the motivation of this work. In Section III, we
present a Markov stochastic model, which explains the relationship
between spectral signatures and community structure. Section IV
describes the critical information derived from the model, such as
stability across multiple timescales and the optimal number of
communities. Our algorithm is formally described in Section V. Then we give some numerical computations for some representative
networks to validate the effectiveness and efficiency of the
algorithm in Section VI . Finally, Section VII concludes this paper.

\section{ Potts model and spin-spin correlation }
The Potts model is one of the most popular models in statistical
mechanics\cite{Wu}. It models an inhomogeneous ferromagnetic
system where each data point is viewed as a marked node
in the network. Here the mark is a cluster label, or spin value, associated
with the node. The configuration of the system is
defined by the interactions between the nodes and controlled by the
temperature. At low temperatures, all labels are identical (spins are aligned),
which is equivalent to the presence of a single cluster. As temperature rises, the
single cluster starts to split and the interactions between
weakly coupled nodes gradually vanished.

Consider an unweighted network with $N$ nodes
without self-loops, a spin configuration $\{S\}$ is defined by
assigning each node $i$ a spin label $s_i$ which may take integer
values $s_i=1,\cdots,q$. Suppose a system of spins can be in
$q$ different states. The Hamiltonian $H(S)$ of a Potts model with
this spin configuration $S$ is given by:

\begin{equation} \label{eq:1}
%\begin{split}
H(S)=\sum_{\langle ij\rangle}J_{ij}(1-\delta_{s_is_j}),   (i,j=1,...,N)
%\end{split}
\end{equation}

where the sum is running over all neighboring nodes denoted as
$\langle ij\rangle$, $J_{ij}$ is the interaction strength between spin $i$ and spin $j$, and $\delta_{s_is_j}$
is 1 if $s_i=s_j$, otherwise 0. $J_{ij}$ is set as
\begin{equation} \label{eq:2}
%\begin{split}
J_{ij}=J_{ji}=\frac{1}{\langle k\rangle}\exp[-\frac{(d_{ij})^2}{2}],   (i,j=1,...,N)
%\end{split}
\end{equation}
where $\langle k\rangle$ is the average number of neighbors per node
and $d_{ij}$ is the Euclidean distance between nodes $i$ and $j$.
The interaction $J_{ij}$ is a monotonous decreasing function of
$d_{ij}$ and the spins $s_i$ and $s_j$ tend to have the same value
as $d_{ij}$ becomes smaller if we minimize the $H(S)$.

To characterize the coherence and correlation between two spins,
spin-spin correlation function $C_{ij}$ is defined as the thermal
average of $\delta_{s_is_j}$\cite{Blatt,Wiseman,Agrawal}:

\begin{equation} \label{eq:3}
%\begin{split}
C_{ij}=\langle \delta_{s_is_j}\rangle
%\end{split}
\end{equation}
It represents the probability that spin variables $s_i$ and $s_j$
have the same value. $C_{ij}$ takes values from the
interval [0, 1], representing the continuum from no coupling to
perfect accordance of spins $i$ and $j$. There are two phases in a
homogeneous system where $J_{ij}$ is determined. At high
temperatures, the system is in the paramagnetic phase and the spins
are in disorder. $C_{ij}\approx\frac{1}{q}$ for all nodes $i$ and
$j$, and $q$ is the number of possible spin values. At low
temperatures, the system turns into the ferromagnetic phase and all
the spins are aligned to the same direction. $C_{ij}\approx1$ holds
for nodes pair $i$ and $j$.

\begin{figure}
\includegraphics[width=8.5cm,height=6.5cm,angle=0]{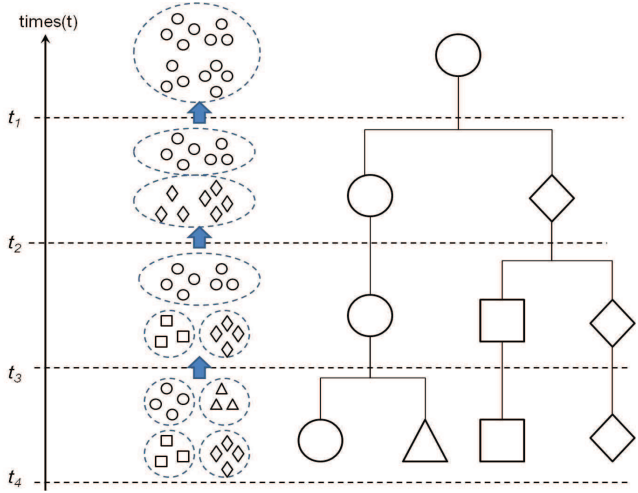} \caption{
Dynamics of spin configuration of four communities ($A,B,C,D$) when
they go through several local uniform states to the global stable
state with temperature decreasing. Different spin values are
described by different shapes. At temperature $t_4$ ($t_4 > t_3 >
t_2 > t_1$, $t_i$ denotes the temperature that $i$ different spin
states in the system), we observe four local uniform spin
state distributions corresponding to four communities. At
temperature $t_3$, the circle and triangle mix together. At $t_2$,
square with diamond mix together in terms of their hierarchical
structure. Finally, at $t_1$, only one spin state is left, in which all nodes have an identical spin distribution.}
\label{fig.1}
\end{figure}

If the system is not homogeneous but has a community structure, the
states are not just ferromagnetic or paramagnetic. We assume that
the spins will go through a hierarchy of local
uniform states (meta-stable states), as shown in
Fig.\ref{fig.1}, before they reach a globally stable state with all the same value
as temperature decreases. In each local uniform state, spin values of nodes within the same communities are
identical and the whole system is divided into several different
local regions (communities) due to the dense
connections. Correspondingly, we can calculate the
hitting and exiting time of each local uniform state to analyze its
stability. The hitting or exiting time is the timescale that the
system just enters or leaves this local uniform state, during which
the nodes' spin values will stably stay on this state. In this way we can
associate the community structure with a local uniform state. For a
well-formed community structure, each community should be cohesive,
which means that it is easy for the nodes to hit the local uniform
state. Thus, the hitting time should be early. At the same time,
communities should stand clear from each other, which means it is
hard for nodes to exit the local uniform state, therefore the
exiting time should be late. Hence, there should be a big gap
between the hitting and exiting times when a well-formed
community structure exists.

Once $J_{ij}$ has been determined, $C_{ij}$ can be obtained by a
Monte Carlo procedure. We used the Swendsen-Wang (SW)
algorithm\cite{Wang} because it exhibits much smaller
autocorrelation time\cite{Wang} than standard methods. For a network with N nodes, the SW algorithm can be briefly described as follows: 1. Generate initial configuration of system $S_1=(s_1,s_2,..,s_N)$ randomly, where $s_i$ is the spin value of node $i$ randomly chosen from 1 to $q$, $q=N/2$ is the initial number of spin values. 2. Generate the configuration of system $S_2$ based on $S_1$: (a) Visit all pair of nodes $<i,j>$ which have interaction $J_{ij}>0$, where $J_{ij}$ is the spin interaction computed only based on the adjacent network. Node $i$ and node $j$ are frozen together with probability:

\begin{equation} \label{eq:3}
%\begin{split}
p_{ij}^f=1-exp(-\frac{J_{ij}}{T}\delta_{s_i,s_j})
%\end{split}
\end{equation}
where $\delta_{s_i,s_j}=1$ if $s_i=s_j$ and 0 otherwise. $T$ is the temperature. Calculate all pairs of spins and put a frozen bond between any frozen pairs. $(b)$ We define SW cluster as the cluster containing all spins that have a path of frozen bonds connecting all of them. Since nodes are frozen only if they have the same spin value, we just need to identify the SW clusters from the same spin values.
$(c)$ For each SW cluster, we draw a random number from $1,2,...,q$ and assign this number to the values of all nodes of this cluster. After going through all SW clusters, the new configuration $S_2$ is generated. 3. Iterate Step 2. Then we can calculate the value $C_{ij}$. We set the initial number of possible spin values $q=N/2$ because if the number of
communities is larger than $q$, some spin states will not be
populated. For a specific node, we choose a initial spin value
randomly from 1 to $q$.

\section{ A stochastic model}
Markov process\cite{Hughes} is a useful tool and has been applied to
find communities\cite{Delvenne,Weinan}. In order to establish
the connection between the community structure and the
local uniform behavior of Potts model, we introduce a Markov
stochastic model featured with spectral signatures for the
network. Let $A =(V,E)$ denote a network, where $V$ is the set of
nodes and $E$ is the set of edges (or links). Consider a Markov
random walk process defined on $A$, in which a random walker freely
walks from one node to another along their links. After arriving at
one node, the walker will randomly select one of its neighbors and
move there. Let $X ={X_t,t\geq 0}$, denote the walker positions, and
$P\{X_t=i,1\leq i\leq N\}$ be the probability that the walker hits
the node $i$ after exact $t$ steps. For $i_t \in V$, we have $P(X_t
= i_t|X_0 = i_0, X_1 = i_1,... ,X_{t-1} = i_{t-1})=
P(X_t=i_t|X_{t-1} = i_{t-1})$. That is, the next state of the walker
is determined only by its current state. Hence, this stochastic
process is a discrete Markov chain and its state space is $V$.
Furthermore, $X_t$ is homogeneous because of $P(X_t = j|X_{t-1} =
i)= p_{ij}$ , where $p_{ij}$ is the transition probability from
node $i$ to node $j$.

To relate the Markov process with the patterns of Potts model,
$p_{ij}$ is defined as

\begin{equation} \label{eq:4}
%\begin{split}
p_{ij}=\frac{C_{ij}}{\sum_{j=1}^NC_{ij}}
%\end{split}
\end{equation}
where $C_{ij}$ is the spin-spin correlation function defined in
Eq.(\ref{eq:3}). Via this representation, the tools of stochastic
theory and finite-state Markov processes
\cite{Delvenne}\cite{Weinan} can be utilized for the purpose of
community structure analysis. Let $P$ be the transition probability
matrix, we have:

\begin{equation} \label{eq:5}
%\begin{split}
P=D^{-1}C
%\end{split}
\end{equation}
where $D$ is the diagonal degree matrix of C. Let $p^{(t)}_{ij}$ be
the probability of hitting node $j$ after $t$ steps starting from
node $i$, we have:

\begin{equation} \label{eq:6}
%\begin{split}
p^{(t)}_{ij}=(P^{t})_{ij}
%\end{split}
\end{equation}

For this ergodic Markov process, $P^t$ corresponds to the
probability of transitions between states over a period of $t$ time
steps. To compute the transition matrix $P^t$, the eigenvalue
decomposition of $P$ is used. If $\lambda_k$ with $k =1,\cdots,N$
denote the eigenvalues of $P$, and its right and left eigenvectors
$u_k$ and $f_k$ are scaled to satisfy the orthonormality
relation\cite{Weinan}:

\begin{equation} \label{eq:7}
%\begin{split}
u_kf_l=\delta_{kl},
%\end{split}
\end{equation}
the spectral representation of $P$ is given by

\begin{equation} \label{eq:8}
%\begin{split}
P=\sum_k\lambda_ku_kf_k
%\end{split}
\end{equation}
and consequently

\begin{equation} \label{eq:9}
%\begin{split}
P^{t}=\sum_k\lambda^{t}_ku_kf_k
%\end{split}
\end{equation}
We assume that eigenvalues of $P$ are sorted such that $\lambda_1 =1
>|\lambda_2|\geq|\lambda_3|\geq ... \geq|\lambda_{N}|$. From the
theory of spectral clustering\cite{Malik,Azran}, $P^{t}$ can
be calculated by a sum of $N$ matrices

\begin{equation} \label{eq:10}
%\begin{split}
P^{t}=\sum_{k=1}^N\lambda^{t}_k\frac{u_nu_n^{T}D}{u_n^{T}Du_n}
%\end{split}
\end{equation}
each of which depends only on $P$'s eigensystem. This is
accomplished by exploiting the fact that $u_n^{T}Du_m=I_{nm}$, because $P$ is
defined by a normalized symmetric correlation
matrix $C$. Because of the largest eigenvalue $\lambda_1=1$, when time
$t\rightarrow \infty$,
$P^{(0)}=P^{\infty}=\frac{u_1u_1^{T}D}{u_1^{T}Du_1}$. The
convergence of every initial distribution to the stationary
distribution $P^{(0)}$ corresponds to the fact that the spin of
whole system ultimately reaches exactly the same value, as
temperature decreases. This perspective belongs to
a timescale $t\rightarrow\infty$, at which all eigenvalues
$\lambda^{t}_k$ go to 0 except for the largest one,
$\lambda^{t}_0=1$. In the other extreme of a timescale $t =0$, $P^t$
becomes the stationary distribution matrix. All of its columns are
different, and the system disintegrates into as many spin values.

The eigensystem of transition matrix $P^t$ can be naturally
correlated with the dynamic process of Potts model. However, it
needs preprocessing due to its
asymmetrical character. We simply extend $P^t$ to the symmetrical
form $G^{(t)}=(P^t+(P^t)^{T})/2$ which is defined as the spin
correlation matrix at time $t$. The eigensystem of $G^{(t)}$ have
the following correlation corresponding to $P^t$:

\begin{lemma}
The eigenvalues and corresponding eigenvectors of matrix $G^{(t)}$ are
exactly same as matrix $P^t$.
\end{lemma}

The proof of lemma 1 is evident. From lemma 1, as $G^{(t)}$ owns the
same eigensystem with $P^t$, it can be used to unfold the dynamic of
Potts states. Also, we can use $G^{(t)}$ to find reasonable
partitions based on many algorithms, such as the K-means algorithm
and GN algorithm\cite{Newman01}.

\section{ Signatures of communities in Potts model across multiple timescales}
In this section, we will uncover the signatures of communities in
Potts model across multiple timescales and use this to identify community structure.
This scheme benefits from the above analysis, namely the connection between Potts model and Markov
process through a stochastic model. A lot of useful information,
such as the optimal number of communities, the stability of networks
at arbitrary timescale, can be uncovered as follows.

Suppose the partition method divides the network $A$ into $K$
clusters or sets $V_k\subset V, k\in{1,2,\cdots, K}$ which are disjoint
and the sets $V_1$, $V_2$,..., $V_K$ together form a partition of
node set $V$. The number of nodes in each cluster is denoted by $N_k=|V_k|$.
Numerically we will deal with the dynamical process
of community structure represented by the spin configuration. We
take the time series into consideration. Therefore, we define
the the signature of a given community $k$ by the ratio of inner
correlations as

\begin{equation} \label{eq:11}
%\begin{split}
S_k^{(t)}=\sum_{i,j\in V_k}\frac{[G^{(t)}]_{i,j}}{N_k}
%\end{split}
\end{equation}
$S_k^{(t)}$ can be viewed as a function of timescale $t$ and we can
use it to study the trend of community structure as time goes on.
Given the number of clusters $K$, the clusters are found by
maximizing the objective function
\begin{equation} \label{eq:12}
%\begin{split}
J_K^{(t)}=\sum_{k=1}^K\sum_{i,j\in V_K}\frac{[G^{(t)}]_{i,j}}{N_K}
%\end{split}
\end{equation}
over all partitions. The objective can be interpreted as the sum of
cluster signature $S_k$ for each cluster $V_k$. The form of
Eq.(\ref{eq:12}) is related to some famous partition measures,
for example, $J_K^{(t)}$ is an extension of the ratio cut criterion
defined as the sum of the number of inter-community edges divided by
the total number of edges through replacing adjacent
matrix $A$ by spin correlation $G^{(t)}$. Furthermore, $J_K^{(t)}$
is also the first part of famous modularity metric $Q$, which is
widely used in the research of community detection.

Further discussion is facilitated by reformulating the average
association objective in matrix form. We denote the membership
vector of node $i$ by $x_i$, a probability vector that describes
node $i$'s involvement in each community. The element $x_i^k$ means
the $k$-th entry of the membership vector of node $i$. The hard
partition and disjointness of sets $V_k$ requires that the
vectors $x_i$ and $x_j$ are orthogonal. The objective
$J_K^{(t)}$ can be written in terms of the indicator vectors $x_k$
as

\begin{equation} \label{eq:13}
%\begin{split}
J_K^{(t)}=\sum_{k=1}^K\frac{x_k^{T} G^{(t)}x_k}{x_k^T x_k}
%\end{split}
\end{equation}

The objective is to be maximized under the conditions $x_k\in
\{0,1\}$ and $x_i^T x_j=0$ if $i\neq j$. Eq.(\ref{eq:8}) can be
rewritten as a matrix trace by accumulating the vectors $u_k$ into a
matrix $X=(x_1, x_2,...,x_K)$. We can then write the objective
$J_K^{(t)}$ as

\begin{equation} \label{eq:14}
%\begin{aligned}
\begin{array}{lcl}
J_K^{(t)}=tr\{(X^T X)^{-1}X^T G^{(t)}X\} \\
=tr\{(X^T X)^{-1/2}X^T G^{(t)}X(X^T X)^{-1/2}\}
\end{array}
%\end{aligned}
\end{equation}
where matrix $X^T X$ is diagonal. The substitution $Y=X(X^T
X)^{-1/2}$ simplifies the optimization problem to $J_K^{(t)}=tr\{
Y^TG^{(t)}Y\}$. The condition $Y^TY=I_K$ is automatically satisfied
since

\begin{equation} \label{eq:15}
%\begin{split}
Y^T Y=(X^T X)^{-1/2}(X^T X)(X^T X)^{-1/2}=I_K.
%\end{split}
\end{equation}

The vectors $y_K$ thus have unit length and are orthogonal to each
other. The optimization problem can be written in terms of the
matrix $Y$ as

\begin{equation} \label{eq:16}
%\begin{split}
\max_{Y^T Y=I}tr\{Y^T G^{(t)}Y\}.
%\end{split}
\end{equation}

\begin{lemma}

\textbf{(Rayleigh-Ritz theorem)} Let $L$ be a symmetric $N\times N$ matrix with eigenvalues $1=\lambda_1\geq
\lambda_2\geq ...\geq \lambda_N$ and the corresponding eigenvactors
$u_1, ... ,u_N$. Then
\begin{equation} \label{eq:17}
%\begin{split}
\max\sum_{k=1}^Ky_k^TLy_k\,\,\,\,\,\,s.t.\,\,\,y_l^Ty_k=I
%\end{split}
\end{equation}
equals $\sum_{k=1}^K \lambda_k$ and the minimum $y_1,...,y_K$ lie in the
subspace spanned by $u_1,...,u_K$.
\end{lemma}

The Rayleigh-Ritz theorem\cite{Azran} tells us that the maximum for
this problem is attained when columns of $Y$ is the eigenvectors
corresponding to the $K$ largest eigenvalues of the symmetric
correlation matrix $G^{(t)}$. We assume that eigenvalues of $P$ are
sorted such that $\lambda_1 =1 >|\lambda_2|\geq|\lambda_3|\geq ...
\geq|\lambda_{N}|$ and the eigenvector corresponding to $\lambda_k$
is denoted as $u_k$. Then the optimal solution of Eq.(\ref{eq:17})
is the matrix $Y=U=\{u_1,...,u_K\}$. And the strength of such a
cluster is equal to its corresponding $t$-th power of the eigenvalue

\begin{equation} \label{eq:18}
%\begin{split}
S_k^{(t)}=\frac{u_k^T G^{(t)}u_k}{u_k^T u_k}=\lambda_k^t\frac{u_k^T
u_k}{u_k^T u_k}=\lambda_k^t
%\end{split}
\end{equation}

For the convergence of the Potts model across multiple timescales,
the vanishing of the smaller eigenvalues as the time growing
describes the loss of different spin states and the removal of the
structural features encoded in the corresponding weaker
eigenvectors. For the purpose of community identification,
intermediate timescales of local uniform states are of interest. If
we want to identify $z$ communities, we expect to find $P^t$ at a
given timescale, the eigenvalues $\lambda^t_k$ may be significantly
different from zero only for the range $k=1,...,z$. This is achieved
by determining $t$ such that $|\lambda_k|^t\thickapprox0$.

From another perspective, because the eigenvalues are sorted by
$\lambda_1 =1 >|\lambda_2|\geq|\lambda_3|\geq ...
\geq|\lambda_{N}|$, the strength of a community at time $t$,
$\lambda_k^t$, can also be viewed as the robustness of $k$-spin
state at time $t$. At this point, the eigengap
$\lambda_k^t-\lambda_{k+1}^t$ can be interpreted as the
``difficulty'' that the $(k+1)$-spin state transfer to the $k$-spin
state at time $t$. Given the correlation matrix $G$, one can measure
the most suitable number of possible spins at a specific time $t$ by
searching for the value $k$ such that the eigengap
$\lambda_k^t-\lambda_{k+1}^t$ is maximized. The number of
communities $\Lambda$ at time $t$ is then inferred from the location
of the maximal eigengap, and this maximal value can be used as a
quality measure for the most stable state. The $\Lambda (t)$ is
formally defined as

\begin{equation} \label{eq:19}
%\begin{split}
\Lambda(t)=arg[max_k(\lambda_k^t-\lambda_{k+1}^t)]
%\end{split}
\end{equation}
From a global perspective if the number of communities $\Lambda$ is
not change for the longest time, we can consider it as the optimal
number for this network, represented as $\Psi$.

%%虽然最优社团数目持续不变，但是内部的变化却不是透明的。
The number of communities $\Lambda$ may keep the same for a long
time. However, the variation of spin configuration hidden behind our
model is still not clear. To reveal the detail of changes, we need
to determine that the timescale of the community structure
represented by spin configuration is robust. To a certain
extent, the most stable state can represent the spin configuration
of the whole network. Thus, we define the stability of community
structure at each timescale, $\Theta(t)$, as the stability of the
most stable spin state:

\begin{equation} \label{eq:20}
%\begin{split}
\Theta(t)=\lambda_{\Lambda(t)}^t-\lambda_{\Lambda(t)+1}^t
%\end{split}
\end{equation}
Our expectation is that from the trend of $\Theta(t)$, one can
find the most stable timescale for community structure where
$\Theta(t)$ reaches the maximal. At this time, the distribution of
spin configuration represents the most suitable community structure.
Furthermore, from a global perspective, we can use the largest
stability corresponding to $q$ communities,
$\Gamma(q)=max\{\Theta(t)|\Lambda(t)=q\}$, to indicate the
robustness of a network, defined as the stability of the structure
with $q$ communities. While $\Gamma(q)$ tries to directly
characterize the network structure rather than a specific network
partition and thus very convenient to estimate the modularity
property of the network.

\begin{figure}
\center
  \subfigure[]{
    \label{fig:subfig:2a} %% label for second subfigure
    \setcounter{subfigure}{1}
    \includegraphics[width=6cm,height=5cm]{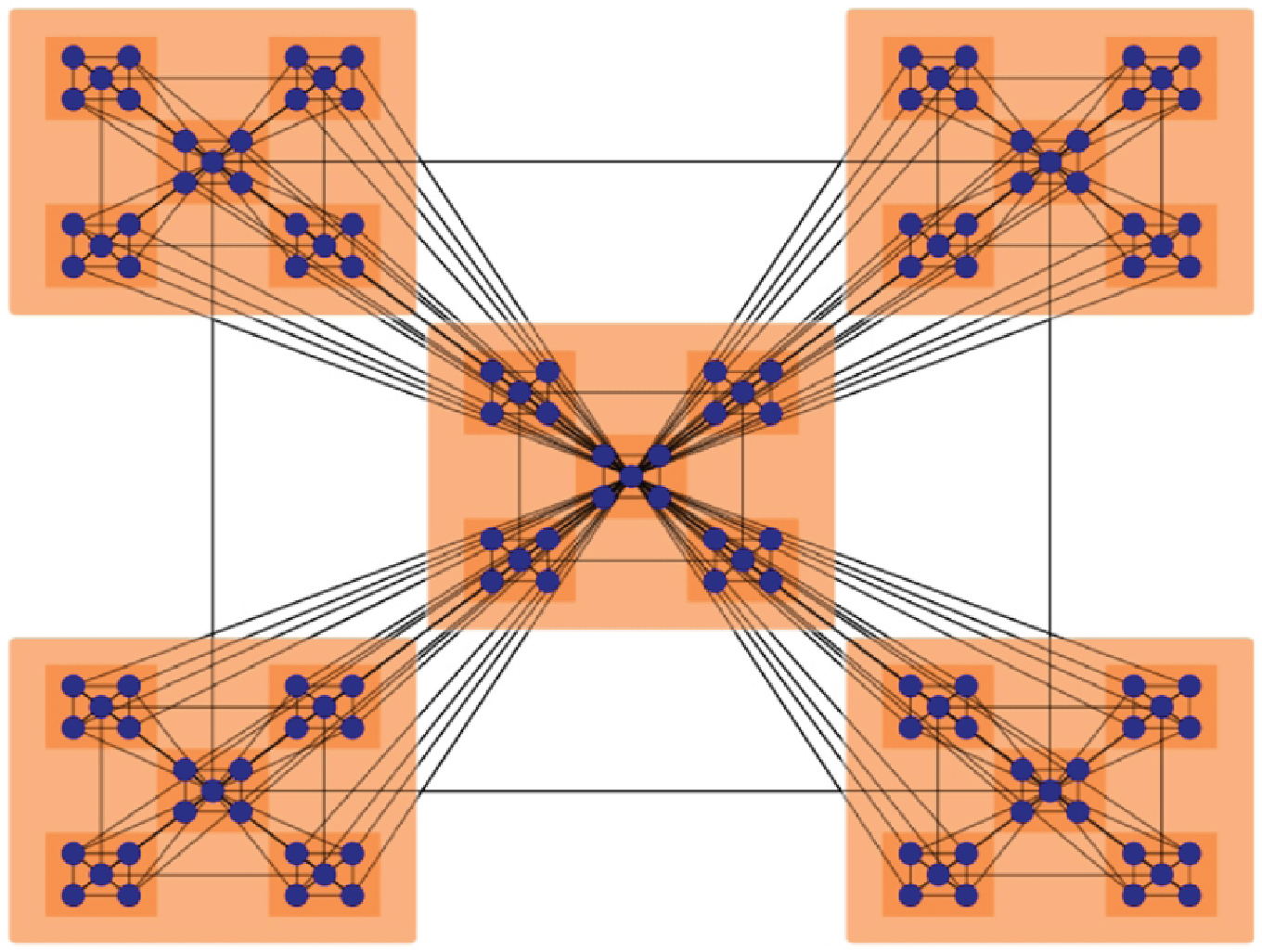}}
  \subfigure[]{
    \label{fig:subfig:2b} %% label for second subfigure
    \setcounter{subfigure}{2}
    \includegraphics[width=8.5cm,height=5.5cm]{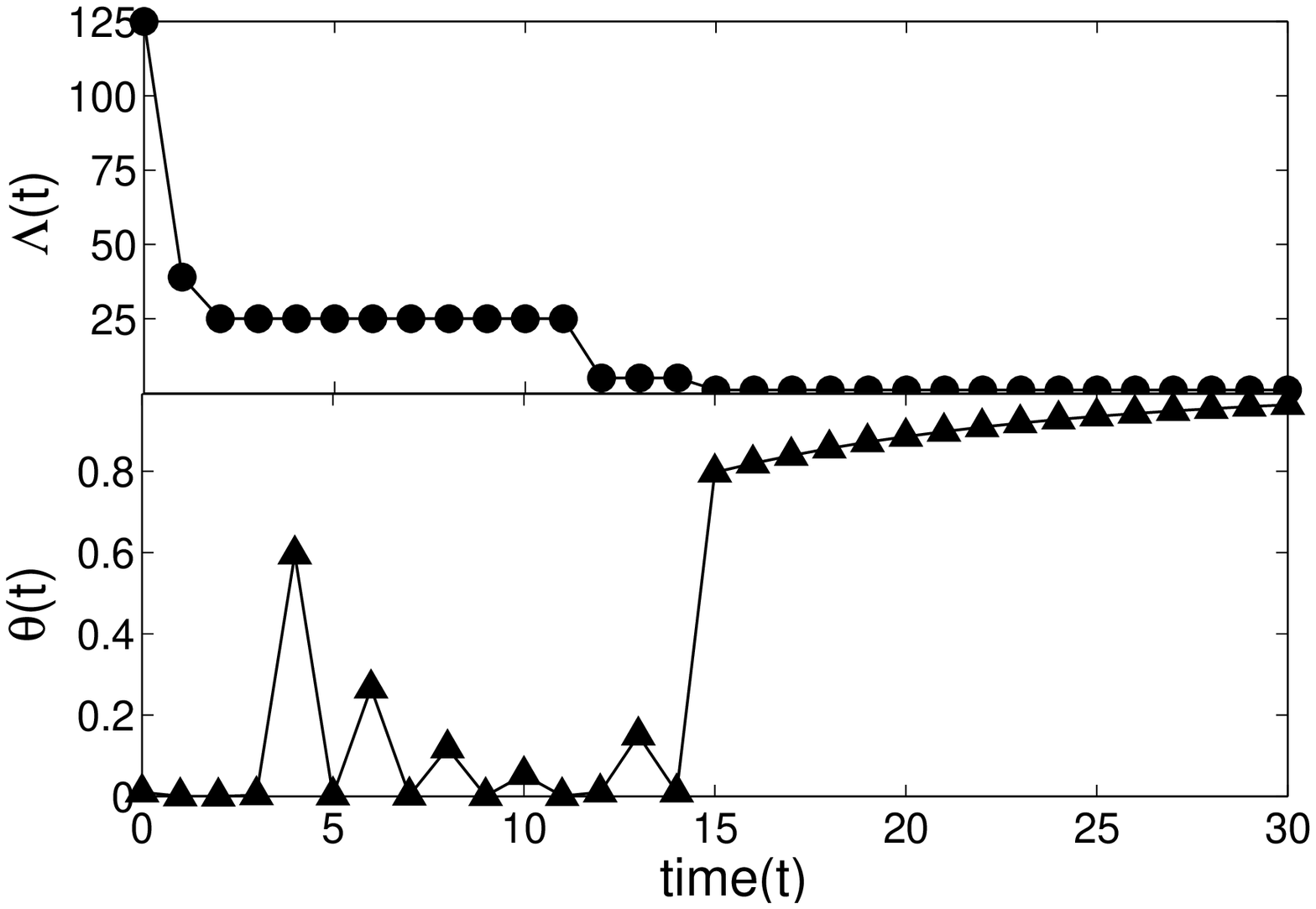}}
\caption{(a) Structure of $RB125$, with 25 dense
communities and 5 sparse communities, are highlighted in the original
network. (b) The value of $\Lambda(t)$ and $\Theta(t)$ versus
time $t$.} \label{fig.2}
\end{figure}

\begin{figure}
\center
  \subfigure[]{
    \label{fig:subfig:3a} %% label for second subfigure
    \setcounter{subfigure}{1}
    \includegraphics[width=6cm,height=5cm]{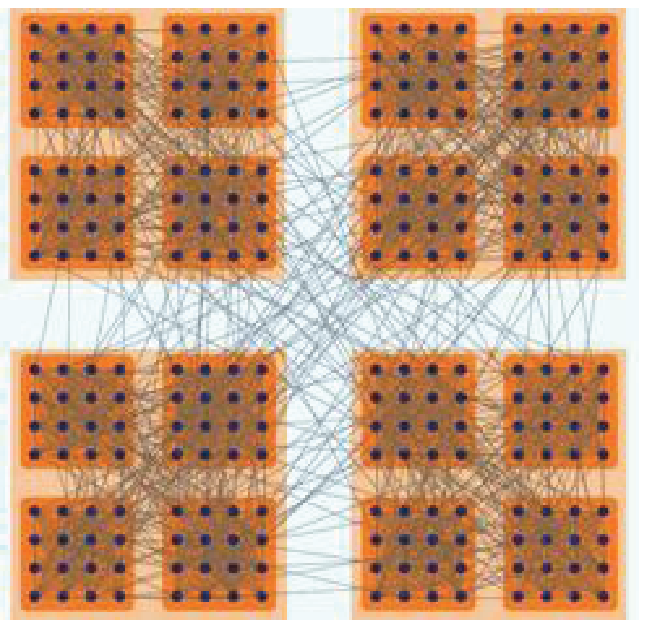}}
  \subfigure[]{
    \label{fig:subfig:3b} %% label for second subfigure
    \setcounter{subfigure}{2}
    \includegraphics[width=8.5cm,height=5.5cm]{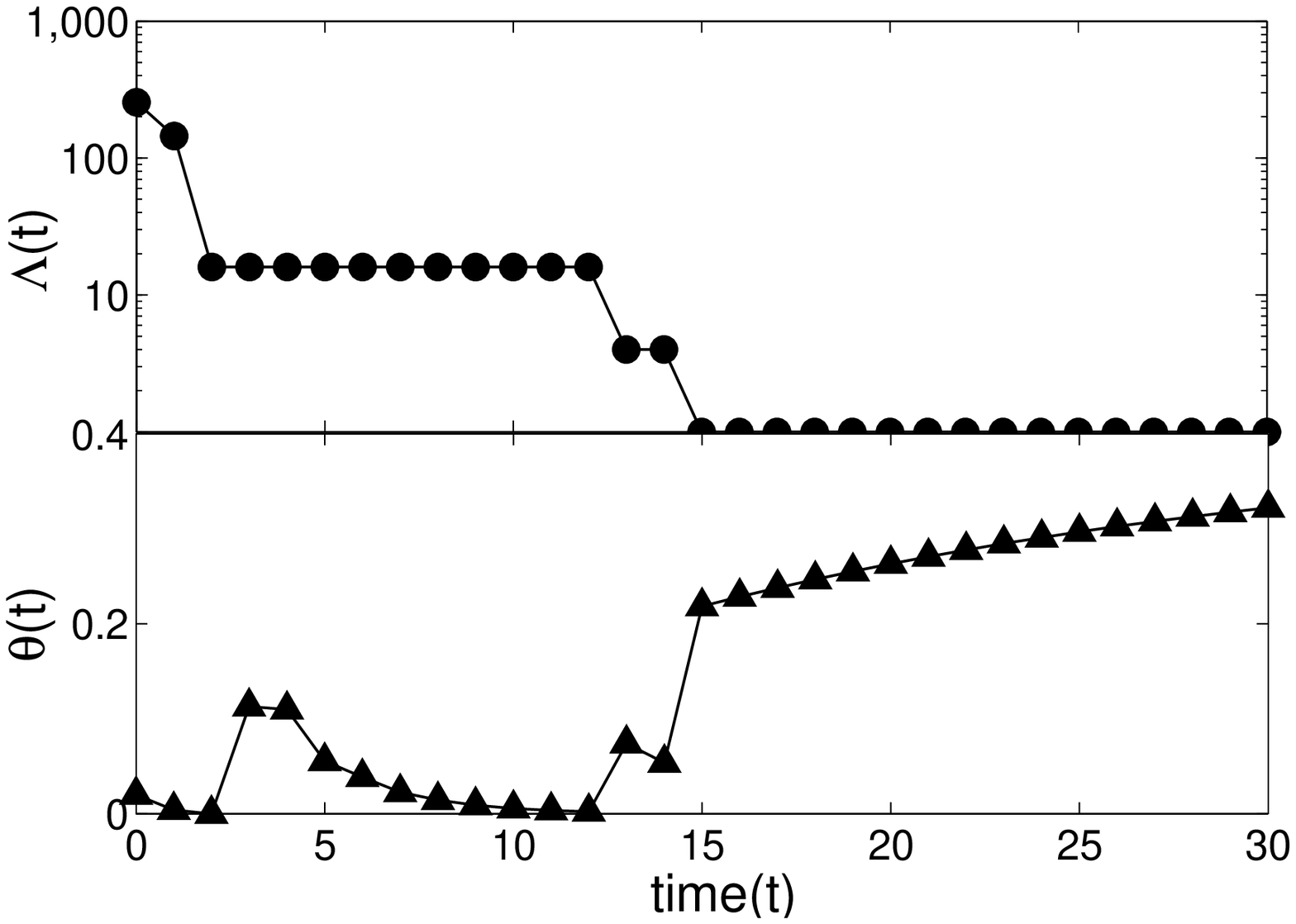}}
\caption{(a) Structure of $H13$-$4$, with 16 dense
communities and 4 sparse communities, are highlighted in the original
network. (b) The value of $\Lambda(t)$ and $\Theta(t)$ versus time $t$.} \label{fig.3}
\end{figure}

To show that the model can uncover hierarchical structures in
different scales, Fig.\ref{fig.2} and Fig.\ref{fig.3} give two
examples of the multi-level community structures.
Fig.\ref{fig:subfig:2a} shows the $RB125$ network, which is
a hierarchical scale-free network proposed by Ravasz and
Barab\'{a}si in \cite{Ravasz}. The regions corresponding to 5 and 25
modules are the most representative in terms of resolution. Next,
$H13$-$4$ proposed by Arenas et al\cite{Arenas} is shown in
Fig.\ref{fig:subfig:3a}, which is a homogeneous degree network
with two predefined hierarchical scales. The first hierarchical
level consists of 4 modules of 64 nodes and the second level
consists of 16 modules of 16 nodes. The partition of both
levels are highlighted on the original networks.

In both examples, the most persistent $\Lambda$ reveals the actual
number of hierarchical levels hidden in a network. The signature of
such levels can be quantified by their corresponding length of
persistent time. The longer the time persists, the more robust the
configuration is. From Fig.\ref{fig:subfig:2b} and
Fig.\ref{fig:subfig:3b}, we can observe 25 and 16 are the optimal
numbers of communities in $RB125$ and $H13$-$4$ networks owning the
longest persistence, respectively. However, 5 modules and 4 modules
are also reasonable partitions which show the fuzzy level of the
hierarchical networks. These results are in perfect consistence
with the generation mechanisms and hierarchical patterns of these two
networks.

Furthermore, we also show that the variation tendency of stability
$\Theta(t)$ in the two cases shed a light on the spin
configuration. From Fig.\ref{fig:subfig:2b} and
Fig.\ref{fig:subfig:3b}, the corresponding stability $\Theta(t)$ is not a parabolic shape for the timescales of a specific $\Lambda$. Thus we cannot easily find the global optimum. However, there are
some local maximal values representing better community
structure. Thus, we can find these local maximal timescales
corresponding to the desirous number of communities and apply $G^t$
to a specific partition method. Furthermore, the stability will reach the lowest value at the end time of all
$\Lambda$. The stability begins to increase when it transits to new status. One can
use $\Theta(t)$ to estimate the modularity property of complex
networks, and the larger the $\Theta$ the stronger the network
community structure. So, one can find the largest corresponding $\Theta$ value for a specific number of community
$\Lambda$ and use it to indicate the robustness of modularity structure. For
$H13$-$4$ shown in Fig.\ref{fig:subfig:3b}, the stability of 16
communities structure, $\Gamma(16)=0.48$ when $t=4$, is larger than
$\Gamma(4)=0.31$ when $t=7$. This indicates that the community
structure containing 16 modules is more robust than community
structure containing 4 modules. Similarly, for $RB125$ network shown
in Fig.\ref{fig:subfig:2b}, $\Gamma(25)=0.48$ corresponding to 25
communities structure when $t=5$ is larger than $\Gamma(5)=0.31$
when $t=7$. The robustness of community structure indicated by
stability $\Gamma(q)$ favors small but obvious modules.
This is the same as \cite{Arenas}\cite{Arenas1} and is reasonable for many
real networks.

\begin{figure}
\center
  \subfigure[]{
    \label{fig:subfig:4a} %% label for second subfigure
    \setcounter{subfigure}{1}
    \includegraphics[width=6cm,height=4.5cm]{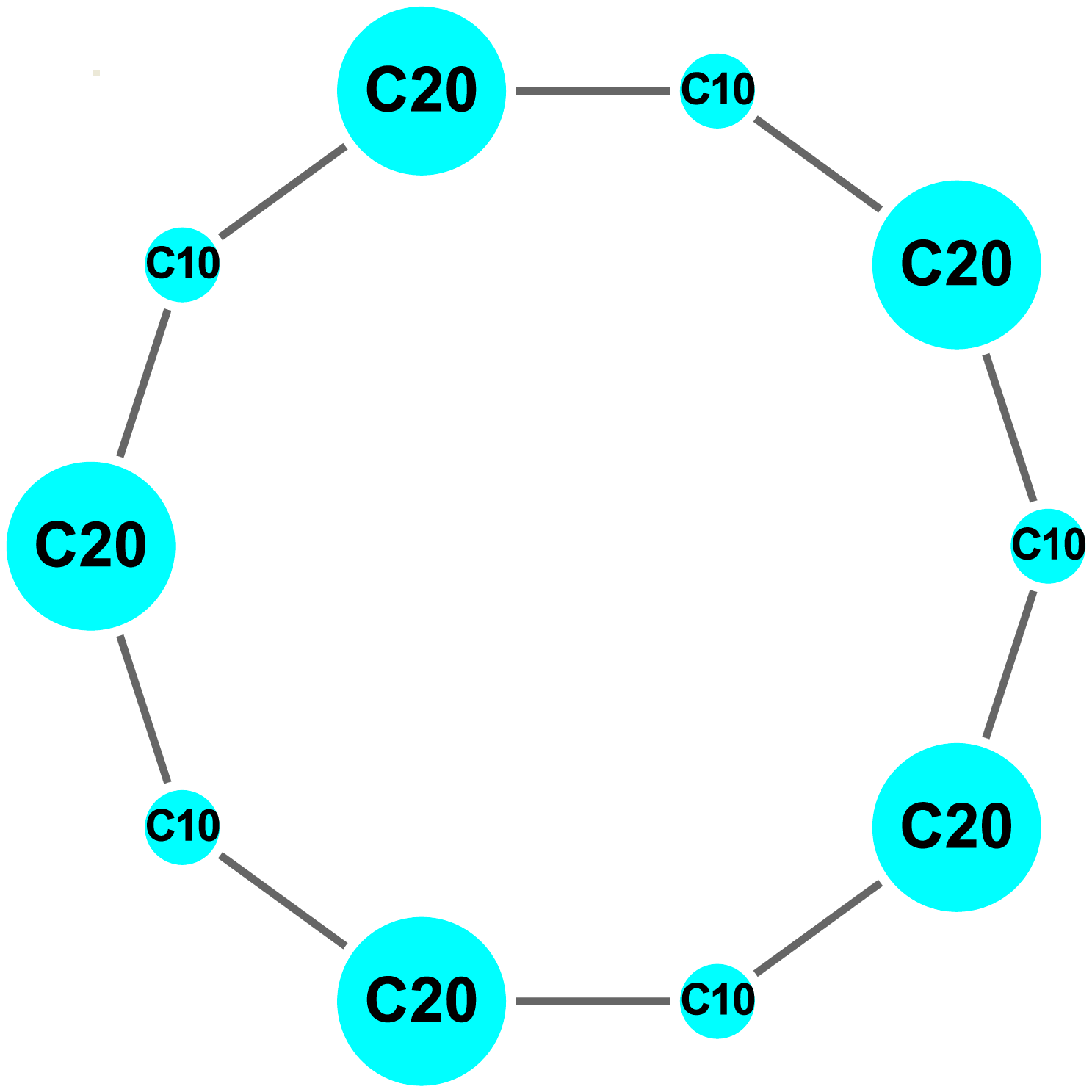}}
  \subfigure[]{
    \label{fig:subfig:4b} %% label for second subfigure
    \setcounter{subfigure}{2}
    \includegraphics[width=8.5cm,height=5.5cm]{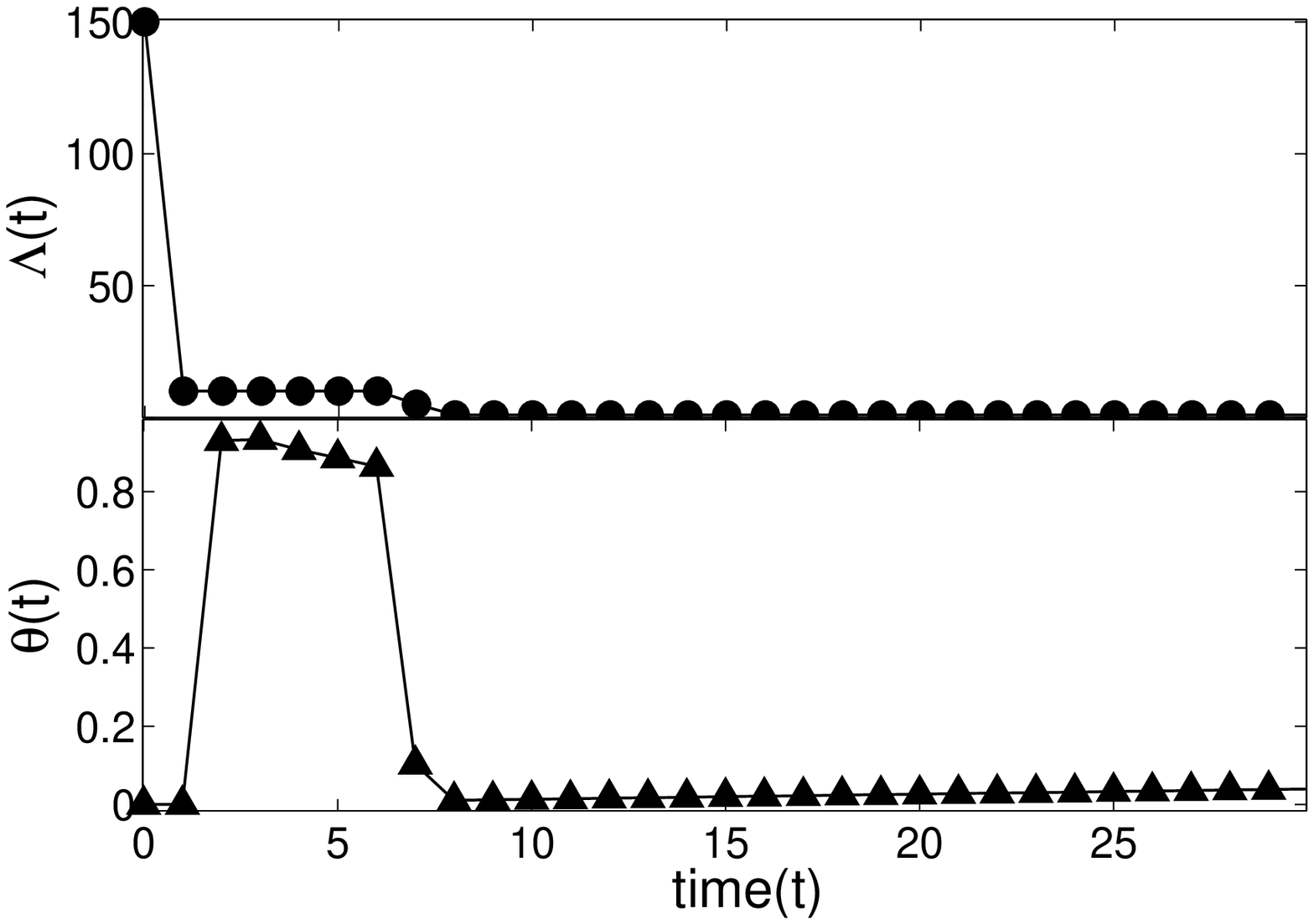}}
\caption{(a) Ring of clique network as a schematic example. Each circle
corresponds to a clique, whose size is marked by its label C20
(contains 20 nodes) or C10 (contains 10 nodes). (b) The value of
$\Lambda(t)$ and $\Theta(t)$ versus time $t$.} \label{fig.4}
\end{figure}

Finally, we emphasize the difference between the stability
measure proposed in this paper and the modularity $Q$ proposed by
Newman\cite{Newman,Newman02}. $Q$ is a well-known
criterion for evaluating a specific partition scheme of a network.
It is defined as ``the fraction of edges that fall within
communities, minus the expected value of the same quantity if edges
fall at random without regard for the community structure''
\cite{Newman02, X,Palla,Li}. Different partition
schemes will get different $Q$ values for the same network, and
larger ones mean better partitions. While our $\Lambda$ and $\Gamma$ try
to directly characterize and evaluate the structure property based on network's spectra, rather than a specific network
partition. Therefore, a network only
has exactly self-deterministic $\Lambda$ and $\Gamma$ values
regardless of how many partition schemes it would have, and the
larger the $\Gamma$ the stronger the network community structure. In
addition, Fortunato \emph{et al}\cite{Fortunato} pointed out the
resolution limit problem of the modularity $Q$, that is, there
exists an intrinsic scale beyond which small qualified communities
cannot be detected by maximizing the modularity. However, as shown
in Fig.\ref{fig.4}, when a clique ring contains cliques with different scales
(i.e.,the heterogeneous community size), the intrinsic community
structure can be exactly revealed by $\Lambda$. With $\Lambda$ and
$\Gamma$, we can quantitatively compare the modularity structure of
different types of complex networks.

\section{A new algorithm to detect community}
To actually perform the community detection, we propose an approach
based on eigenvalue decomposition\cite{Allefeld} of correlation
matrix $G^{(t)}$. This algorithm allows us to identify multivariate communities
across multiple timescales. Based on the above analysis,
we correlate the multivariate community structure with the
dynamics of the eigenvalues and eigenvectors.

The eigenvalues $\lambda_k$ and eigenvectors $u_k$ of the symmetric
and real-valued matrix $G^{(t)}$ can be obtained by solving the
eigenvalue equation

\begin{equation} \label{eq:21}
%\begin{split}
G^{(t)}\cdot u_k=\lambda_k^t\cdot u_k,   k=1,...,N
%\end{split}
\end{equation}

which has $N$ different solutions when time $t$ is
small. Assume that the eigenvectors are
normalized ($\sum_i u_k(i)=1$). Each signature $S_k(t)=\lambda_k^t$
is associated with a specific community (the elements in the vector have the
same spin value) and quantifies its strength at a given timescale.
For each community $k$, the internal structure is described by the
corresponding eigenvector $u_k$. After normalization ($\sum_i
u_k(i)=1$), its components quantify the relative involvement of each
node $i$ to community $k$ by $u_k^2(i)$. Combining the signature of
the community and the index $u_k^2(i)$, the ``absolute'' involvement
of node $i$ in a community $k$ at time $t$ can be described by the
following participation index,

\begin{equation} \label{eq:22}
%\begin{split}
R^{(t)}_k(i)=\lambda_k^tu_k^2(i)
%\end{split}
\end{equation}
Node $i$ is considered as belonging to community $k$
when the participation index becomes maximal.

From Eq.(\ref{eq:22}), we observe that
participation index evolves as time goes on. When the timescale
$t\rightarrow \infty$, all eigenvalues $\lambda^{t}_k$ approach to 0
except the largest one, $\lambda^{t}_0=1$. At this time, all
nodes belong to the same community according to the participation
index definition. In the other extreme when $t =0$, the participation
matrix $R$ actually becomes the eigenvector matrix $U^2$. All of its
columns are different, and the number of
communities is equal to the dimension of the matrix. Here we are interested in
the optimal partition at an intermediate timescale with large
stability $\Theta(t)$, when the spin configuration represents the most
robust community structure. So, we first determine the optimal
number of communities by using $\Lambda$ across long time $t$.
Then, we pick up the timescale $t$ that the
stability $\Theta(t)$ is maximal between and $\Lambda(t)$ equals to the
optimal number of communities. In many real networks, the formulation of communities is a hard partition and each node belongs
to only one cluster after the cluster. This is often too restrictive
for the reason that nodes at the boundary among communities share commonalities with more than
one community and play a role of transition in many diffusive networks. In our work,
the participation index $R$ motivates the extension of the partition to a probabilistic setting.
It is extended to the fuzzy partition concept where each node maybe long to different communities
with nonzero probabilities at the same time and more reasonable for the real world.  Finally, we calculate the
participation index at the most stable time $t$. The framework
of the whole process is summarized in Algorithm \ref{alg:Framwork}. In the process of the algorithm, calculate the spin-spin correlation matrix $C$
is based on $SW$ algorithm and costs less than $O(N^2)$. It is easy to estimate the computational cost of the algorithm is main on the calculation of eigensystem of $G$ and for sparse graphs, it is about $O(N^2)$. Other steps of the process are some simple matrix computations. So finally, we obtain the cost of Algorithm \ref{alg:Framwork} is $O(N^2)$. Our algorithm is a parameter free method and very easy to implement in real networks.

\begin{algorithm}[htb] %算法的开始

\caption{ Framework of our new algorithm.} %算法的标题

\label{alg:Framwork} %给算法一个标签，这样方便在文中对算法的引用

\begin{algorithmic}[1] %这个1 表示每一行都显示数字

\REQUIRE ~~\ %算法的输入参数：Input

The adjacent matrix of the network $A$;\

\ENSURE ~~\ %算法的输出：Output

%Ensemble of classifiers on the current batch, $E_n$;

\STATE Calculate the spin-spin correlation matrix $C$.

\STATE Calculate the Markov transition probability matrix $P$ and $G$ based on $C$.

\STATE Calculate the eigenvalues and corresponding eigenvectors of $G$.

\STATE Find the optimal number of communities $K$ and corresponding
times $t$ with the largest stability.

\STATE Calculate the participation index $R$ according to Eq.(\ref{eq:22}).

\RETURN Output the participation index $R$; %算法的返回值

\end{algorithmic}

\end{algorithm}

\section{ Experiments}
In this section, we will benchmark the performance of our algorithm. We
designed and implemented three experiments for two main purposes:
(1) to evaluate the accuracy of the algorithm; (2) to apply it to
real large-scale networks.

\subsection{Benchmark network}

\begin{figure}
 \centering
\subfigure[]{
    \label{fig:subfig:5a} %% label for second subfigure
    \setcounter{subfigure}{1}
    \includegraphics[width=7cm,height=5cm]{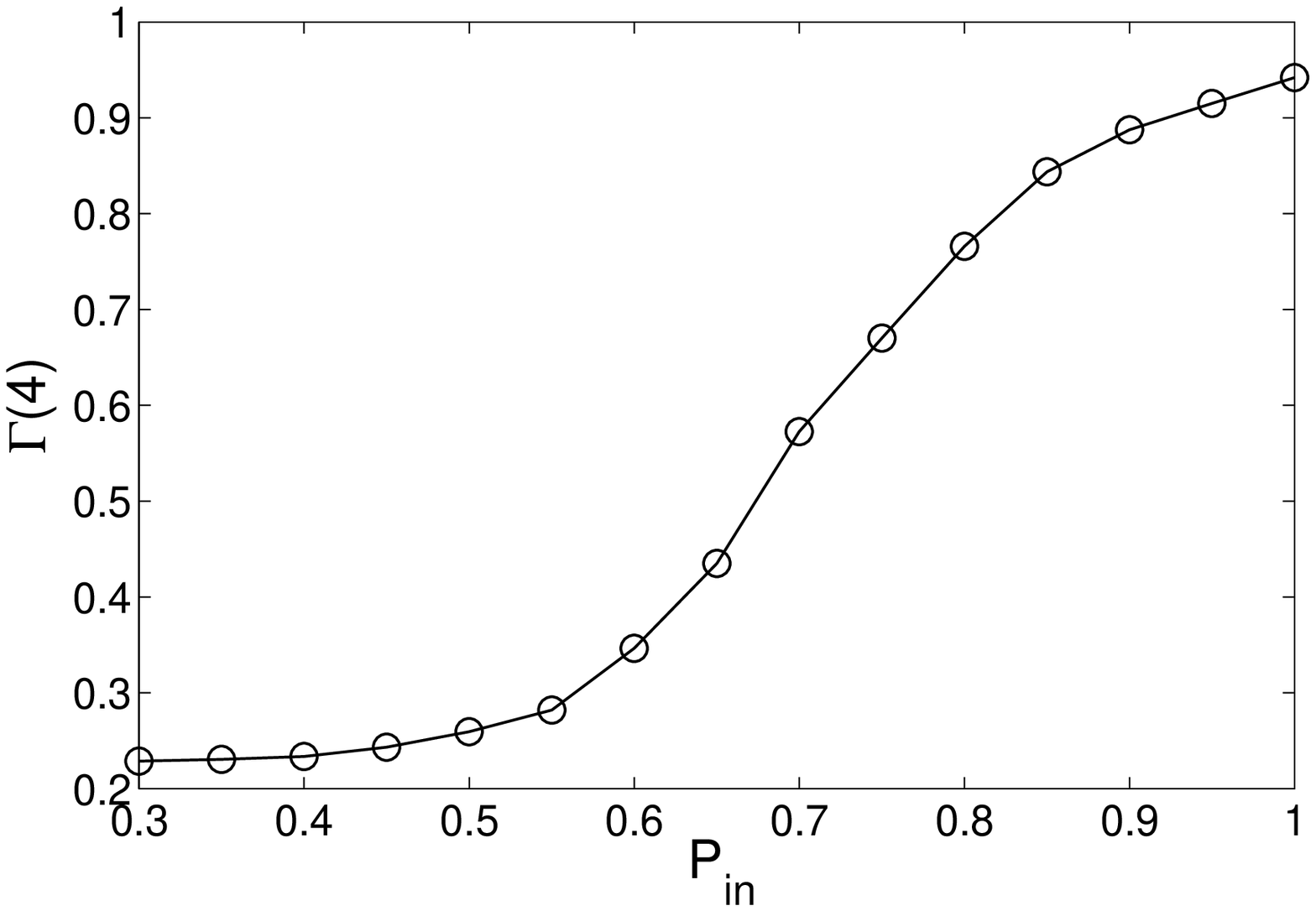}}
  \subfigure[]{
    \label{fig:subfig:5b} %% label for second subfigure
    \setcounter{subfigure}{2}
    \includegraphics[width=8.5cm,height=5cm]{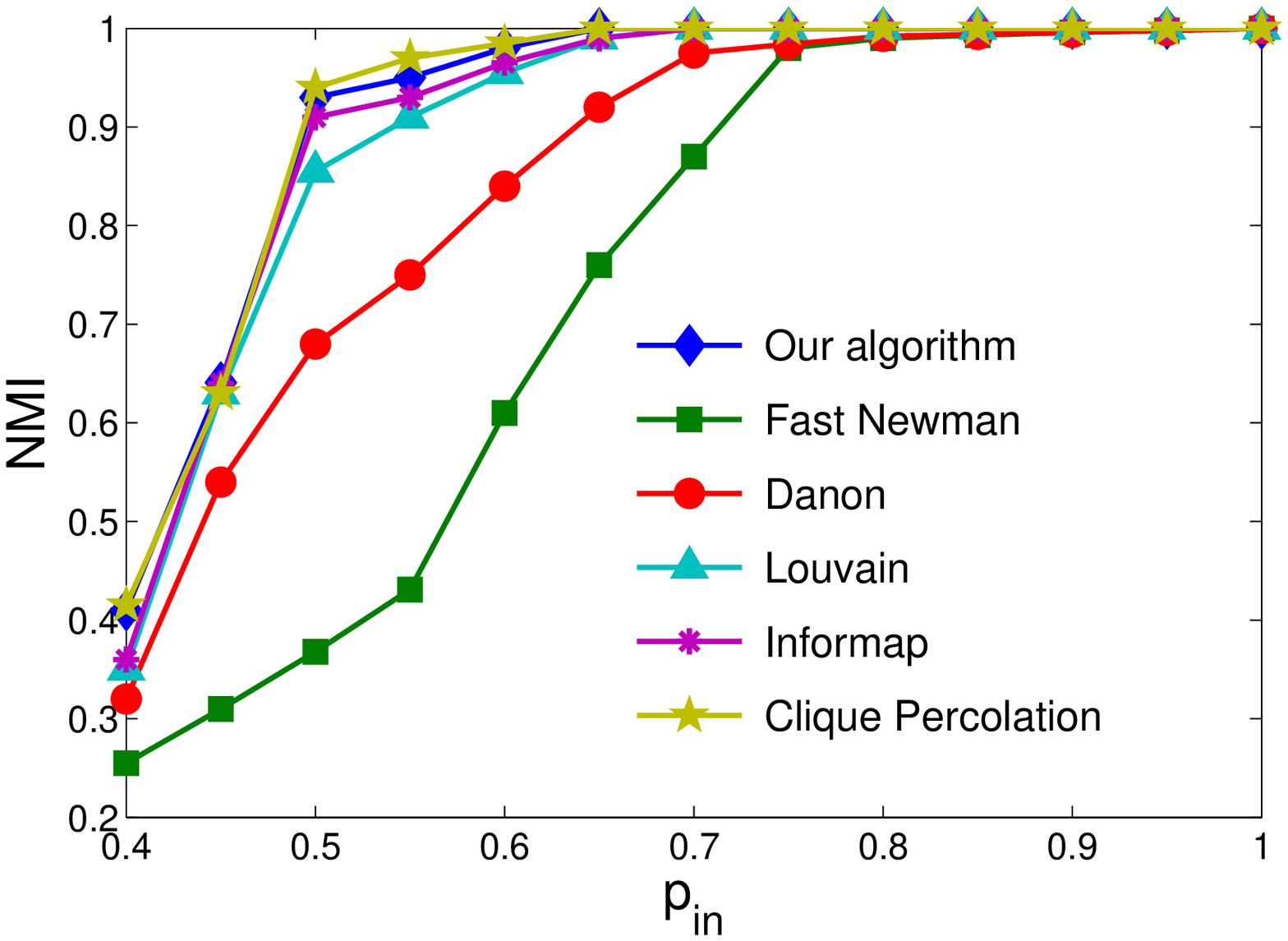}}
\caption{(a) $\Gamma(4)$ values of networks versus different
$P_{in}$. (b) Comparison of accuracy of our algorithm with other five existing algorithms.}
\label{fig.5}
\end{figure}

We empirically demonstrate the effectiveness of our algorithm
through comparison with other five well-known algorithms on the
artificial benchmark networks. These algorithms include: Newman's
fast algorithm\cite{Newman}, Danon et al.'s method\cite{Danon}, the
Louvain method\cite{Blondel}, Infomap\cite{Rosvall}, and the clique
percolation method\cite{Palla}. We utilize widely used Ad-Hoc
network model, which can produce a randomly synthetic network
containing 4 predefined communities and each has 32 nodes. The
average degree of nodes is 16, and the ratio of intra-community
links is denoted as $P_{in}$. As $P_{in}$ decreases, the community
structures of Ad-Hoc networks become more and more ambiguous, and
correspondingly, their $\Gamma(4)$ values climb from 0 to 1, as
shown in Fig.\ref{fig:subfig:5a}.

We use the normalized
mutual information (NMI) measure\cite{Lancichinetti01} to qualify
the partition found by each algorithm. We ask the question whether
the intrinsic scale can be correctly uncovered. The experimental
results are illustrated in Fig.\ref{fig:subfig:5b}, where y-axis
represents NMI value, and each point in curves is obtained by averaging the values
obtained on 50 synthetic networks. As we
can see, all algorithms work well when $1-\mu$ is more than 0.7
with NMI larger than $0.85$. Compared with other five algorithms,
our algorithm performs the best. Its accuracy is only slightly
worse than that of the clique percolation when
$0.5\leq1-\mu \leq 0.65$. However, the complexity of the clique
percolation is more than $O(n^3)$ and nearly the same as the time
consuming Breadth First Search($BFS$). By contrast, the time
complexity of our method is very low($O(n^2)$) and can be easily implemented.

\subsection{ US Football network }
The United States college football team network has been widely used
as a benchmark example\cite{Newman}\cite{Li}
due to its natural community structure. We used the data gathered by Girvan and Newman\cite{Newman}. It is a
representation of the schedule of Division I American Football games
in the 2000 season in USA. The nodes in the network
represent the 115 teams, while the edges represent 613 games played
in the course of the year. The whole network can be naturally divided into 12
distinct groups. As a result, games are generally more frequent between
members of the same group than between members of different groups.

\begin{figure}
\includegraphics[width=9cm,height=5.5cm,angle=0]{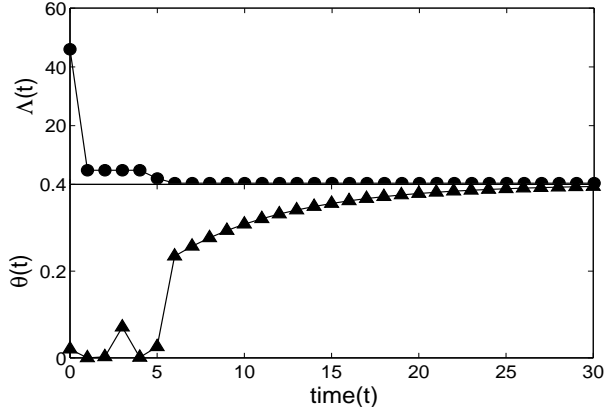} \caption{
Computational results of $\Lambda(t)$ and $\Theta(t)$ on US football
network.} \label{fig.6}
\end{figure}

\begin{figure*}
\includegraphics[width=12cm,height=8.5cm,angle=0]{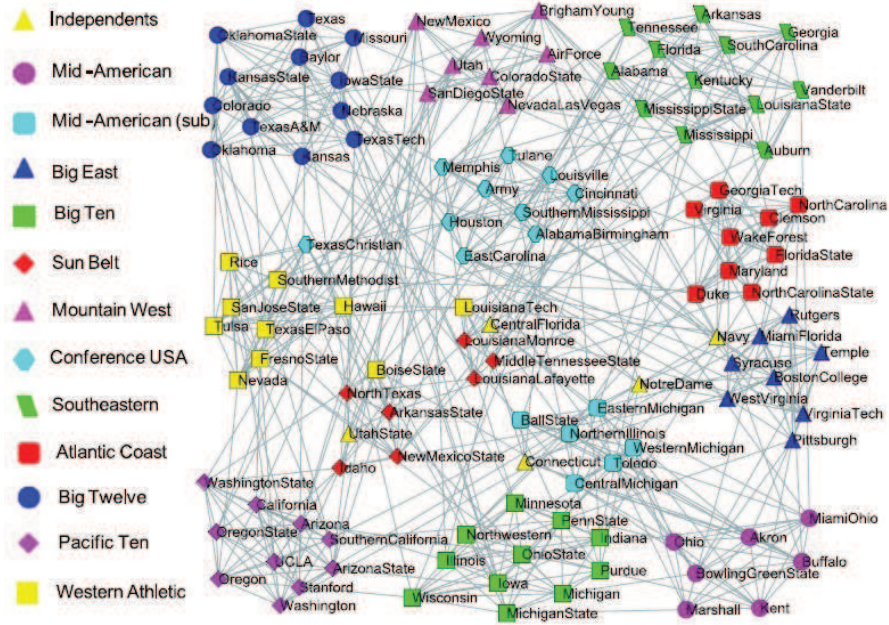} \caption{
Computational results of our algorithm on the football team network.
The nodes with the same shapes and colors are teams in the same
group, and the dense subgraphs in the layout are communities
detected by the algorithm. Four fuzzy overlapping nodes are described as independents.} \label{fig.7}
\end{figure*}

First, we calculate $\Lambda$ and the corresponding stability
$\theta$ and the results are illustrated in Fig.\ref{fig.6}. Results show that the optimal
number of communities is $\Lambda=12$, which perfectly agree with the true situation.
The stability $\theta$ reaches $\Gamma(12)=0.31$ when $t=4$. Then we
apply our algorithm to the football team network and partitions the
network into 12 communities, which is shown in Fig.\ref{fig.7}. The
correct rate of our method is more than $93\%$, which means that the
detected community structure is in a high agreement with the true
community structure. Actually, methods based on optimization of
modularity $Q$ usually can just find 11 communities and the correct
rate is low due to the fuzziness of the network. We concludes that the ability of our
method to reveal a natural characteristic is valuable for many real
networks. Furthermore, our algorithm has identified 5 interesting overlapping nodes
which described as yellow triangles. The nodes are all fuzzily lie at the boundary communities
and can be viewed as some relative independent clubs which can be interpreted readily by the human eye.

\subsection{ Scientific collaboration network }

\begin{figure}
\centering
 \subfigure[]{
    \label{fig:subfig:8a} %% label for first subfigure
    \setcounter{subfigure}{1}
    \includegraphics[width=5cm,height=5cm]{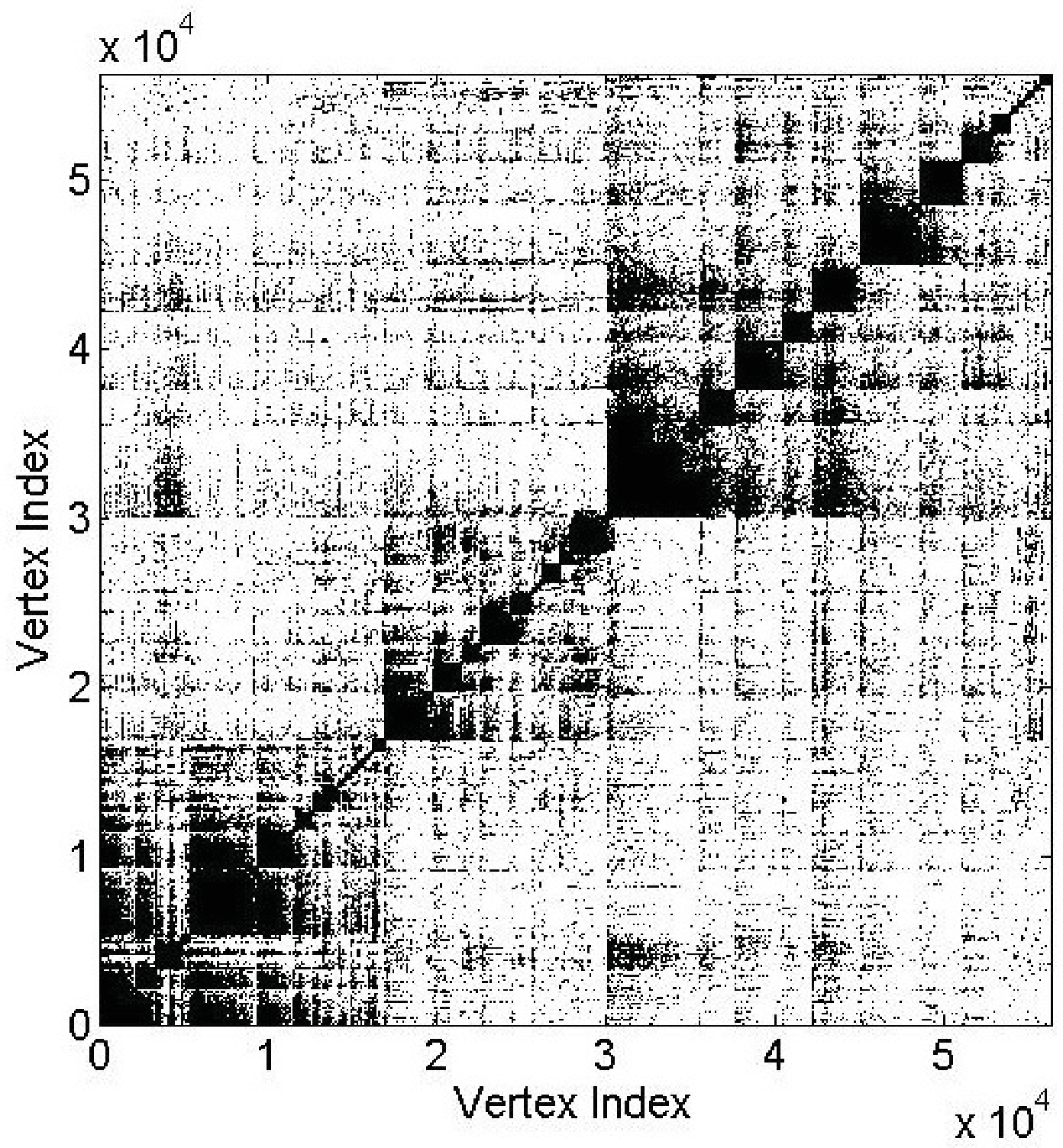}}
  %\hspace{0.5in}
  \subfigure[]{
    \label{fig:subfig:8b} %% label for second subfigure
    \setcounter{subfigure}{2}
    \includegraphics[width=6.5cm,height=5cm]{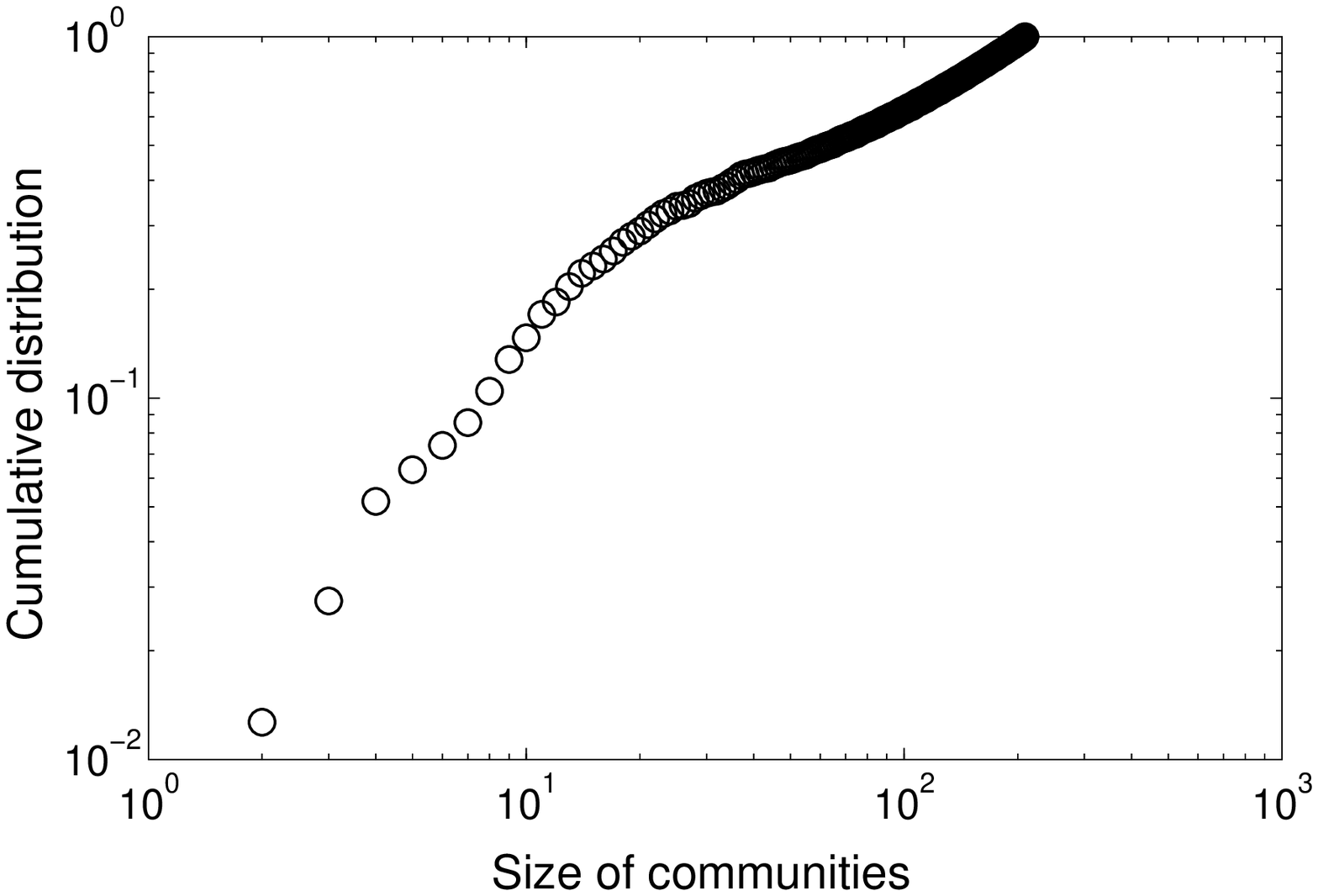}}
  \subfigure[]{
    \label{fig:subfig:8c} %% label for second subfigure
    \setcounter{subfigure}{3}
    \includegraphics[width=10cm,height=5.5cm]{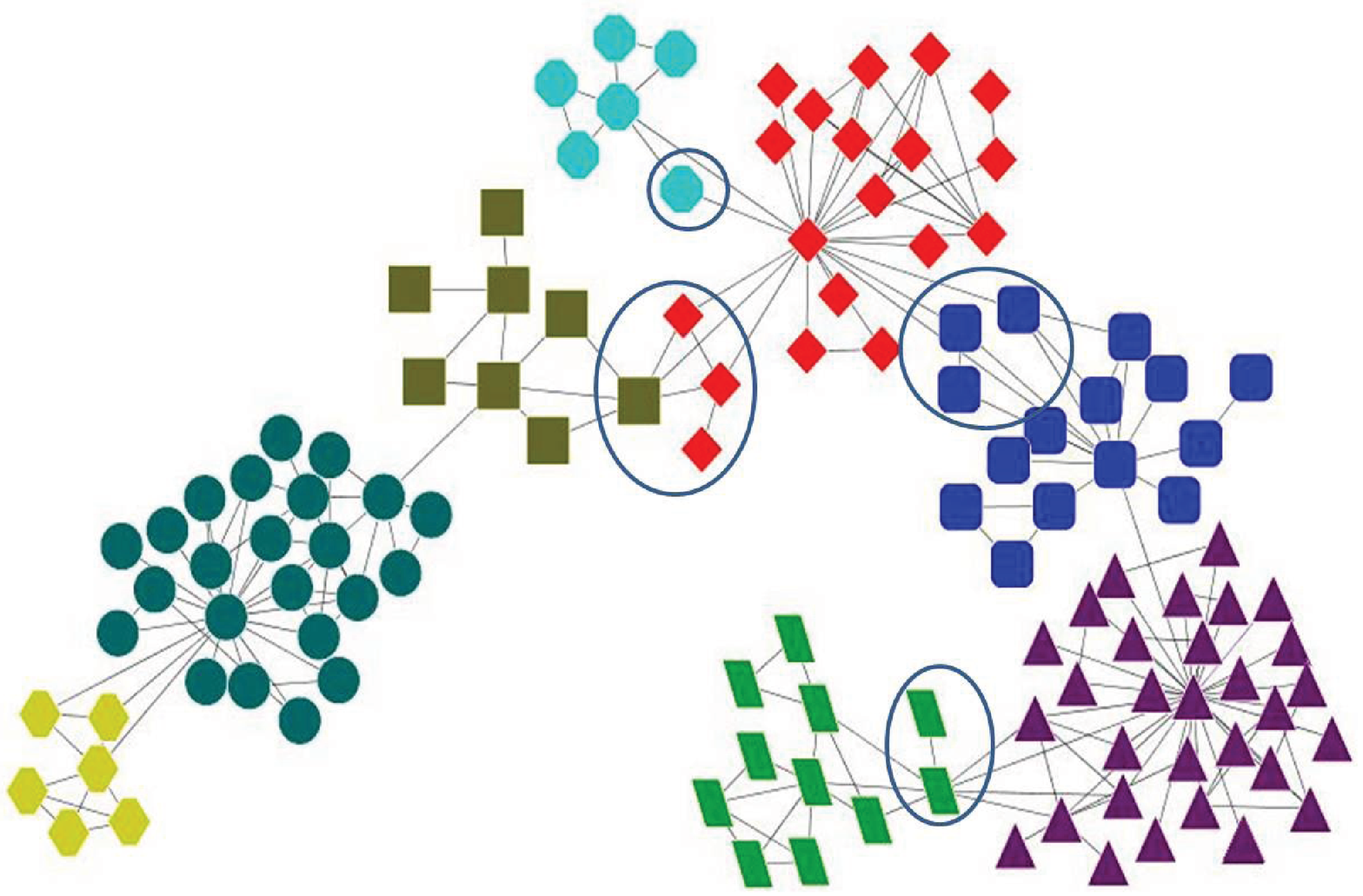}}
\caption{(a) Transformed adjacency matrix of the scientific collaboration network.
  (b) Distribution of community sizes in a linear plot.
  (c) Subnetwork including eight communities illustrated in different shapes and colors and 10 overlapping nodes enclosed by four circles.}
\label{fig.8}
\end{figure}

Finally we tested our algorithm on a large-scale network, the
scientific collaboration network, collected by Girvan and Newman
\cite{Girvan}. The network illustrates the research collaborations
among 56,276 physicists in terms of their coauthored papers posted on
the Physics E-print Archive at \url{arxiv.org}. Totally, this network
contains 315,810 weighted edges. For visualization purpose, our
algorithm outputs a transformed adjacency matrix (in which the nodes
within the same community are grouped together) with a
hierarchical community structure. From the transformed matrix of
Figs.\ref{fig:subfig:8a}, one can observe a quite strong community
structure, or a group-oriented collaboration pattern. Among these
physicists, three biggest research communities are self
organized regarding to three main research fields: condensed matter,
high-energy physics (including theory, phenomenology and nuclear),
and astrophysics.

The cumulative distribution of community sizes in power plot is
shown in Fig.\ref{fig:subfig:8b} and it is a typical scale-free distribution
which exists broadly in real world. In total, 737 communities were
detected by the optimal community stability, the maximum size of
those communities is 195, the minimum size is 2, and the average
size is 76. Among these communities, 1,433 of 6,931 pairs of
communities have fuzzy participation index with each other. $5\%$
largest communities contain $25.4\%$ of the nodes, while the
others are relatively small. The three largest communities
correspond closely to research subareas. The largest is
solid-state physics, the second largest is super-nuclear
physics, and the third is theoretical astrophysics. Furthermore, a
subnetwork including eight communities in is shown in Fig.\ref{fig:subfig:8c}
and four regions including 10 overlapping nodes are highlighted by four circles, which were detected according to the
participation index $R$. The partition result is completely the same as the
results in Refs.\cite{Girvan} and \cite{Li}. The efficient performance in large
real network indicates that our method is useful for further researches in various
fields.

\section{ Conclusion}

In summary, we have presented a more theoretically-based community detection framework
which is able to uncover the connection between network's community structures
and spectrum properties of Potts model's local uniform state.
We demonstrate that important information related to
community structures can be mined from a network's spectral
signatures through a Markov process computation, such as the stability of modularity structures and the
optimal number of communities. Based on theoretical analysis, we
further developed an algorithm to
detect fuzzy community structure. Its effectiveness and
efficiency have been demonstrated and verified through both the
simulated networks and the real large-scale networks.

%%%%%%%%%%%%%%%%%%%%%%%%%%%%%%%%%%%%%%%%%%%%%%%%%%%%%%%%%%%
\vskip 1mm \vspace{0.3cm}

\noindent{\bf Acknowledgments:} We are grateful to the anonymous reviewers for their valuable suggestions which are very helpful for improving the manuscript. The authors are separately supported by NSFC grants 11131009, 60970091, 61171007, 91029301, 61072149, 31100949, 61134013 and grants kjcx-yw-s7 and KSCX2-EW-R-01 from CAS.

\end{document}